\documentstyle[aps,12pt]{revtex}
\begin{document}
\draft
\title
{Relativistic Mean Field Approach and the Pseudo-Spin Symmetry}
\author{G. A. Lalazissis$^1$, Y. K. Gambhir$^{1,2}$, J. P. Maharana$^2$, 
C. S. Warke$^2$, and P. Ring$^1$}
\address{
$^1$Physik-Department der Technischen Universit\"at M\"unchen,
D-85747 Garching, Germany \\
$^2$ Department of Physics, Indian Institute of Technology Bombay\\
Powai, Mumbai 400026, India} 
\maketitle
\centerline{\today}
\begin{abstract}
Based on the Relativistic Mean Field (RMF) approach the existence of the
broken pseudo-spin symmetry is investigated. Both spherical RMF and
constrained deformed RMF calculations are carried out employing realistic
Lagrangian parameters for spherical and for deformed sample nuclei. The
quasi - degenerate pseudo-spin doublets are confirmed to exist near the fermi
surface for both spherical and deformed nuclei.
\end{abstract}

\pacs{PACS numbers : 21..60Cs, 21.10.-k, 24.10.Jv}


\vspace{1cm}

Pseudospin symmetry has been discovered in nuclear physics nearly 30 years
ago \cite{HA.69,AHS.69,RDH.73}. The recent claim \cite{Gin.97} that
pseudo-spin symmetry may arise due to near equality in magnitude of
attractive scalar and repulsive vector fields in relativistic mean theory,
has revived the activity related to the understanding of the origin of this
symmetry in real nuclei. The concept of pseudo-spin symmetry \cite
{HA.69,AHS.69} is based on the experimental observation of the existence of
quasi-degenerate doublets of normal parity orbitals ($n$, $\ell $, $j=\ell +%
\frac{1}{2}$) and ($n-1$, $\ell +2$, $j=\ell +\frac{3}{2}$) such as (4$%
s_{1/2}$, 3$d_{3/2}$), (3$d_{5/2}$, 2$g_{7/2}$) etc., in the same major
shell. Since for spherical systems the quantum numbers $j^{\pi }$ are
conserved, the pseudo-spin angular momenta ($\tilde{\ell}$, $\tilde{s}=1/2$)
satisfy $\tilde{j}=j=\tilde{l}\pm \frac{1}{2}$.

In order to interpret this near degenerate pair of $j=\ell +1/2$ and $j=\ell
+3/2$ states as pseudo-spin doublets corresponding to $\tilde{m}_{s}=\pm 1/2$%
, $\tilde{\ell}$ has to be $\ell +1$. It then follows for the major
oscillator quantum number: $\tilde{N}$ = $N-1$, for the radial quantum
number $\tilde{n}=(\tilde{N}-\tilde{\ell})-1$ and for the parity 
$\tilde{\pi}=-\pi $. For zero
pseudo-spin orbit splitting, the pseudo-spin multiplet will be degenerate.
Thus the pair of orbitals (4$s_{1/2}$, 3$d_{3/2}$) and (3$d_{5/2}$, 2$g_{7/2}
$) can be viewed as the (2$\tilde{p}_{1/2}$, 2$\tilde{p}_{3/2}$) and (1$%
\tilde{f}_{5/2}$, 1$\tilde{f}_{7/2}$) pseudo-spin doublets. The symmetry can
also be investigated in deformed nuclei. In the asymptotic Nilsson scheme
one finds the pseudo-spin quantum numbers ($\tilde{N}=N-1$ , $\tilde{n}%
_{3}=n_{3}$ , $\tilde{\Lambda}=\Lambda +1$ and $\tilde{\Omega}=\Omega $) .
Therefore, the Nilsson orbitals $[N,n_{3},\Lambda ,\Omega =\Lambda +1/2]$
and $[N,n_{3},\Lambda +2,\Omega =\Lambda +3/2]$ can be viewed as the pseudo
spin-orbit doublets $[\tilde{N},\tilde{n}_{3},\tilde{\Lambda},\tilde{\Omega}=%
\tilde{\Lambda}\pm 1/2]$ \cite{BMH.82}.

Apart from the rather formal relabeling of quantum numbers various proposals
for an explicit transformation from the normal scheme to the pseudo-spin
scheme have been made in the last twenty years and several nuclear
properties have been investigated in this scheme \cite
{Mot.91,BDM.92,CMQ.92,BBD.96}. However, the origin of pseudo-spin symmetry
remained unknown until the recent observation of Ginocchio 
\cite{Gin.97,GL.97}
where for the first time the origin of this symmetry is claimed to be
revealed as due to the near equality in magnitude of the attractive scalar
and repulsive vector fields in relativistic theories. Here in this letter we
follow this idea and investigate to what extent the pseudo-spin symmetry
is broken for realistic cases. For this purpose we concentrate as well on
spherical as on deformed nuclei and we use the framework of relativistic
mean field (RMF) theory\cite{SW.86}. It has been shown that this
phenomenological approach is very successful in describing the ground state
nuclear properties of spherical, deformed and also for nuclei far away from
the beta stability line (see for example\cite{Rei.89,GRT.90,Rin.96}.

The RMF starts with a Lagrangian density describing the nucleons as Dirac
spinors $\psi $, of mass $m$, interacting via the meson ($\sigma $-, $\omega 
$-, and $\rho $-) and the electromagnetic fields. The standard Lagrangian
density used in the RMF theory is written as\cite{GRT.90}:\newline
\begin{eqnarray}
{\cal L} &=&\bar{\psi}\left( \gamma (i\partial -g_{\omega }\omega -g_{\rho }%
\vec{\rho}\vec{\tau}-eA)-m-g_{\sigma }\sigma \right) \psi   \nonumber \\
&&+\frac{1}{2}(\partial \sigma )^{2}-U(\sigma )-\frac{1}{4}\Omega _{\mu \nu
}\Omega ^{\mu \nu }+\frac{1}{2}m_{\omega }^{2}\omega ^{2}  \nonumber \\
&&-\frac{1}{4}{\vec{{\rm R}}}_{\mu \nu }{\vec{{\rm R}}}^{\mu \nu }+\frac{1}{2%
}m_{\rho }^{2}\vec{\rho}^{\,2}-\frac{1}{4}{\rm F}_{\mu \nu }{\rm F}^{\mu \nu
}  \label{E1}
\end{eqnarray}
It includes a nonlinear self-interaction $U(\sigma )$ of the $\sigma $%
-field:\newline
\begin{equation}
U(\sigma )=\frac{1}{2}m_{\sigma }^{2}\sigma ^{2}+\frac{1}{3}g_{2}\sigma ^{3}+%
\frac{1}{4}g_{3}\sigma ^{4}.  \label{E2}
\end{equation}
which takes into account in a phenomenological way the density dependence of
the parameters of the model. $m_{\sigma }(g_{\sigma })$, $m_{\omega
}(g_{\omega })$, $m_{\rho }(g_{\rho })$ are the respective meson masses
(coupling constants) and $g_{2}$ and $g_{3}$ are the coupling strengths of
the nonlinear sigma field $U(\sigma )$.

It is straightforward to write the coupled baryon spinor and the mesons
mean field equations. Starting from the Dirac equation for the single
nucleon radial wave function with the spherical attractive scalar ($%
S=-g_{\sigma }\sigma $) and the repulsive vector ($V=g_{\omega }\omega $)
potentials and following the standard procedure, by eliminating the small
components ($g_{i}$), the large components ($f_{i}$) obey the following
second order differential equation: 
\begin{eqnarray}
&&\left\{ -\nabla ^{2}-\frac{S^{\prime }+V^{\prime }}{2m-E-(S+V)}\left( 
\frac{\partial }{\partial r}+\frac{\kappa _{i}+1}{r}\right) \right\} \,f_{i}
\nonumber \\
&=&-\left( 2m-E-(S+V)\right) (E-(S-V))\,f_{i}.  \label{E3}
\end{eqnarray}

Here the eigenvalues denoted by $\kappa _{i}$, of the operator $-\beta ({\bf %
\Sigma \cdot L}+1)$ are given by 
\begin{equation}
\kappa _{i}~=~\mp \left( j_{i}+{\frac{1}{2}}\right) \qquad {\rm for}\qquad
j_{i}=\ell _{i}\pm {\frac{1}{2}}\ ,  \label{E4}
\end{equation}
and $S^{\prime }$ ($V^{\prime }$) are the derivatives of the potentials $S$ (%
$V$ ) with respect to $r.$ The binding energy $E\ge 0$ is measured with
respect to the nucleon mass $M$ in natural units $\hbar =c=1$.

On the other hand eliminating the large component $f_{i}$ we have for the
small component $g_{i}$ the following second order differential equation: 
\begin{eqnarray}
&&\left\{ -\nabla ^{2}-\frac{S^{\prime }-V^{\prime }}{E-(S-V)}\left( \frac{%
\partial }{\partial r}-\frac{\kappa _{i}-1}{r}\right) \right\} \,g_{i} 
\nonumber \\
&=&\left( 2m-E-(S+V)\right) (E-(S-V))\,g_{i}.  \label{E5}
\end{eqnarray}
For the case of equal strengths, $S=V$, the Eq.\ (\ref{E5}) reduces to: 
\begin{equation}
-\nabla ^{2}g_{i}+E(S+V)\,g_{i}=E(2m-E)\,g_{i}.  \label{E6}
\end{equation}
Clearly Eq.\ (\ref{E6}) has an energy dependent potential ($E(V+S)$) and has
the eigenvalue $E(2m-E)$. After scaling the radial variable $r=x/(\sqrt{E})$%
, the potential has a complicated $(\sqrt{E})$ dependence {\it i.e.,} $%
S\left( {x/\sqrt{E}}\right) +V\left( {x/\sqrt{E}}\right) $. In such a
situation this equation~(\ref{E6}) is no longer a normal Schr\"{o}dinger
eigenvalue equation. Further, it is obvious that in this equation all
solutions with ``bound'' states in the Fermi sea with $E\geq 0$ are shifted
to one degenerate eigenvalue with $E=0$, which, in fact, is not bound. The
corresponding wave functions are not normalizable. This indeed is an
unphysical situation. This equation is the same as the equation (3) of Ref.~%
\cite{Gin.97} in the scaled variable $x$ when written in terms of the
partial waves and using the relation $\ell (\ell +1)=\kappa (\kappa -1)$.
Here $\ell $, the angular momentum of the lower component $g_{i}$ is
identified with the pseudo-spin angular momentum ($\tilde{\ell}$). This is
the pseudo-spin symmetry limit of Ref.~\cite{Gin.97}, where the doublets $%
j~=~\tilde{\ell}\pm 1/2$ with the same $\tilde{\ell}$ are degenerate.
However, in this limit only the Dirac sea states exist and no Dirac valence
bound states and therefore contradicts reality. According to these
considerations in all realistic situations the pseudo-spin symmetry must be
broken. Therefore the question arises, to which extent it is broken in real
nuclei. So far only the spherical case has been investigated for square well
potentials \cite{Gin.97} and for spherical solutions of the RMF equations%
\cite{GM.97,MST.97}.

In the present letter we investigate the broken pseudo-spin symmetry both
for the spherical and deformed nuclei within the relativistic mean field
approach. For our study, we choose $^{208}$Pb as a representative of
spherical nuclei and $^{154}$Dy as a representative of deformed nuclei. We
use in our calculations the Lagrangian parameter set NL3\cite{LKR.97} which
successfully reproduces the ground state properties of nuclei, spread over
the entire periodic table. The other parameter sets like NL1 and NLSH (see 
\cite{Rin.96}) are expected to give almost identical results for these
nuclei.

First, spherical RMF calculations in the coordinate space are carried out
for $^{208}$Pb. The calculated binding energy and the charge radius agree
remarkably well with the experiment. The calculated single particle energies
for the bound orbitals near the fermi surface are shown in Fig.~(\ref{figA}a)
for
neutrons and protons. It is clear from the figure that the pairs of bound
neutron valence orbitals (2$g_{7/2}$,3$d_{5/2}$) and (1$i_{11/2}$,2$g_{9/2}$%
) which correspond to pseudo-spin doublets (2$\tilde{f}_{7/2}$,2$\tilde{f}%
_{5/2}$) and (1$\tilde{h}_{11/2}$,1$\tilde{h}_{9/2}$) respectively, are
quasi-degenerate indicating only a small breaking of pseudo-spin symmetry.
The same is more or less true for the pairs of neutron hole ((2$f_{5/2}$,3$%
p_{3/2}$),(1$h_{9/2}$,2$f_{7/2}$)), proton valence (particle) (1$h_{9/2}$,2$%
f_{7/2}$)), and proton hole ((2$d_{3/2}$,3$s_{1/2}$), (1$g_{7/2}$,2$d_{5/2}$%
)), orbitals forming the pseudo-spin doublets. But here the energy
separation between the partners of the respective doublets is relatively
larger. The larger is the binding energy the larger is the separation. This
indicates that the concept of the pseudo-spin symmetry becomes better and
better for the orbitals as their energies approach closer and closer to the
continuum. This is consistent with the results found in Ref.\cite{Gin.97}
for the square well potentials. In addition, the energy separation becomes
larger, if the pseudo-orbital angular momentum ($\tilde{\ell}$) increases.
The dependence of the energy splitting of the pseudo-spin partners on the
energy E and on the pseudo-orbital angular momentum $\tilde{\ell}$ can
easily be understood from Eq.(\ref{E5}). For a given pseudo-orbital angular
momentum $\tilde{\ell}$ the term in Eq.(\ref{E5}) which splits the
pseudo-spin partners is: 
\begin{equation}
\frac{S^{\prime }-V^{\prime }}{(S-V)-E}\,\,\frac{\kappa _{i}}{r}
\end{equation}
It has the energy dependence $(E-(S-V))$ in the denominator. Now $(S-V)$ is
about 50 MeV. Bound states in the Fermi sea have a binding energy $E<50$
MeV. For increasing binding energy $E$ , i.e. going to more deeply bound
states, the denominator decreases. This then results in a larger energy
splitting between the pseudo-spin partners. For example for the orbit $%
\tilde{\ell}=3$ the energy splitting between the pseudo-spin partners (1$%
g_{7/2}$ and 2$d_{5/2}$)) will be relatively larger as compared to that
between (2$g_{7/2}$ and 3$d_{5/2}$)). In addition, the bigger is the value
of $\tilde{\ell}$ the larger is the splitting. For instance,the energy
splitting between the pseudo-spin partners (1$i_{11/2}$ and 2$g_{9/2}$)
corresponding to $\tilde{\ell}=5$ is relatively larger as compared to that
between the partners (2$g_{7/2}$ and 3$d_{5/2}$) which corresponds to $%
\tilde{\ell}=3$, in the same major shell. Interestingly, the sign of the
energy splittings between the partners of the neutron valence doublets is
opposite to those of the neutron hole, proton particle and proton hole
doublets.

The normalized single nucleon wave functions (both large ($f$) and small 
($g$) components) are plotted for the pseudo-spin partners corresponding to the
valence neutron pairs, the neutron hole pairs and valence proton pairs in
Figs.~\ref{figA} (b), (c) and (d) respectively. The phase of the lower 
components ($g$%
) of one of the partners is reversed while plotting, in order to exemplify
the differences in the lower components of the pseudo-spin partners.
Clearly, the lower components are much smaller in magnitude as expected and
are almost equal in magnitude. In the case of exact pseudo-spin symmetry,
the lower component of the pseudo-spin partners should be identical (except
for the phase). The very small differences between these $g$'s which mainly
appear around the surface are negligible for the pseudo-spin partners having
very small binding energies.

Next we consider deformed systems and impose constraint on the quadrupole
moment. Constrained relativistic Hartree calculations have been carried out
for the nucleus $^{154}$Dy. The coupled differential equations for the
spinors and the meson fields are given in Ref. \cite{GRT.90}. They have been
solved by expanding the spinors and the meson fields in terms of anisotropic
oscillator wave functions. Numerical details are given in Refs. \cite{GRT.90}
and \cite{RGL.97}. Pairing correlations are treated in the constant gap
approximation and the Lagrangian parameter set NL3\cite{LKR.97} is used. The
calculated potential energy surface is shown in Fig.~\ref{figB}. 
The value of the
calculated ground state deformation parameter $\beta _{2}$ is 0.202 which is
to be compared with 0.237, the corresponding experimental value. The
calculated ground state binding energy 1262.95 MeV differs from the
corresponding experimental value by merely 1.2 MeV.\newline
The energies of the bound neutron pairs of orbitals corresponding to
pseudo-spin doublets are plotted against the deformation $\beta _{2}$
ranging from 0.0 to 0.5 in Fig.~\ref{figC}. 
The asymptotic Nilsson quantum numbers $%
[N,n_{3},\Lambda ,\Omega ]$ are good for large values of the deformation $%
\beta _{2}$. The pseudo-spin doublets $[\tilde{N},\tilde{n}_{3},\tilde{%
\Lambda},\tilde{\Omega}=\tilde{\Lambda}\pm 1/2]$ \cite{BMH.82} are indicated
by $[\tilde{N},\tilde{n}_{3},\tilde{\Lambda}]$ 
$\uparrow $ and $\downarrow $
in the figure. For zero deformation ($\beta _{2}=0$) the orbitals are
indicated by the corresponding spherical states. The figure reveals the
following:\vspace{0.2cm}

\begin{itemize}
\item[(i)]  The energy splitting between the pseudo-spin partners is smaller
for the valence orbitals and for the partners just below the Fermi surface.

\item[(ii)]  This energy difference is relatively larger for the partners
having larger pseudo-spin angular momentum ($\tilde{\ell}$).

\item[(iii)]  In general, this separation stays almost constant and does not
vary with deformation after reasonable value of $\beta _{2}$.

\item[(iv)]  The energy difference between the $\downarrow $ and the $%
\uparrow $ partners always remains positive except for $[\tilde{{\bf {404}}}]
$, where there is crossing at around $\beta =0.3$. Such a crossing is not
very unusual, it has also been observed in Ref.\cite{BMH.82}.\vspace{0.2cm}
\end{itemize}

These systematics are consistent with those observed in the spherical case
above. A similar plot for the proton pseudo-spin doublets shown in 
Fig.~\ref{figD}
reveals identical systematics as those observed for the neutron case
(Fig.~\ref{figC}). It is interesting to note that in Ref.\cite{BMH.82} the energy
difference between the valence neutron pseudo-spin partners is negative
(opposite to ours) while it has the same sign as ours for protons. This may
be due to the negative value obtained for ${V}{\it _{\ell s}}$, the strength
of the pseudo-spin orbit interaction, from the Nilsson parameterization for $%
82<N<126$.

Similar calculations have also been carried out for other spherical and
deformed nuclei and they show identical systematics. The conclusions
presented here, are therefore rather general.

In conclusion, it is shown in the relativistic mean field framework that
quasi-degenerate pseudo-spin doublets do exist near the fermi surface for
both spherical and deformed nuclei. The pseudo-spin symmetry is restored
better and better as one moves closer to the continuum limit. These
conclusions confirm the findings of Ginocchio \cite{Gin.97,GL.97}.
\vspace{1cm}

One of the authors (G.A.L) acknowledges support from the DAAD. The work is
also supported in part by the Bundesministerium f\"{u}r Bildung und
Forschung under the project 06 TM 875.


\begin{figure}
\caption{Pseudo-spin splitting in the spherical nucleus $^{208}$Pb:
(a) single particle spectra in the vicinity of the Fermi surface for
neutrons $\nu$ and protons ($\pi$) and large ($f$) and small ($g$)
components of the Dirac wave functions for the pseudo-spin doublets
$\nu2\tilde d$ (b), $\nu2\tilde f$ (c) and  $\pi2\tilde g$ (d)}
\label{figA}
\end{figure}

\begin{figure}
\caption{Energy surface of the deformed nucleus $^{154}$Dy as function
of the quadrupole moment $q$ in units of $barn$}
\label{figB}
\end{figure}

\begin{figure}
\caption{Single particle energies of the deformed Dirac equation for the
neutrons in the nucleus $^{154}$Dy as a function of the 
quadrupole deformation parameter 
$\beta_2$. Asymptotic pseudo-spin quantum numbers are given and the
pseudo-spin partners are indicated by arrows 
$\uparrow $ and $\downarrow $}
\label{figC}
\end{figure}

\begin{figure}
\caption{Single particle energies for protons in $^{154}$Dy, for details
see Fig.~\ref{figC} }
\label{figD}
\end{figure}

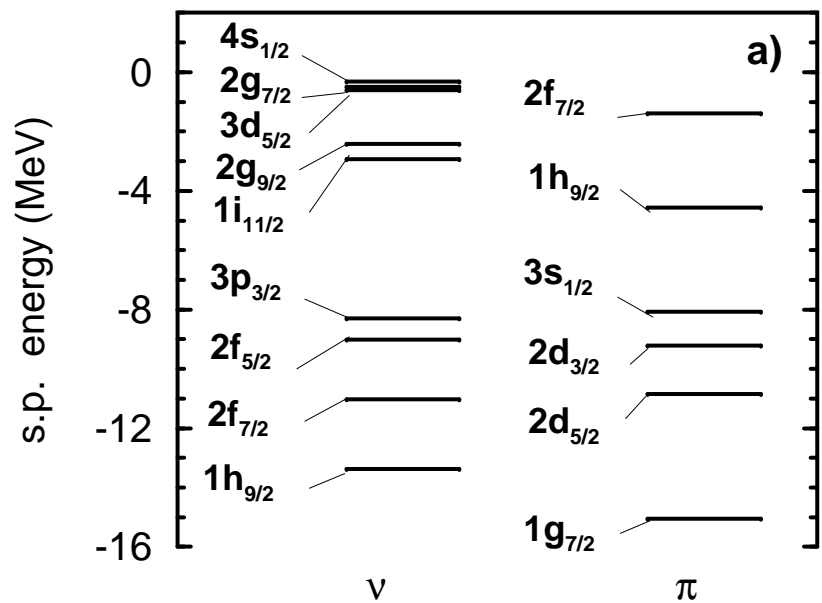
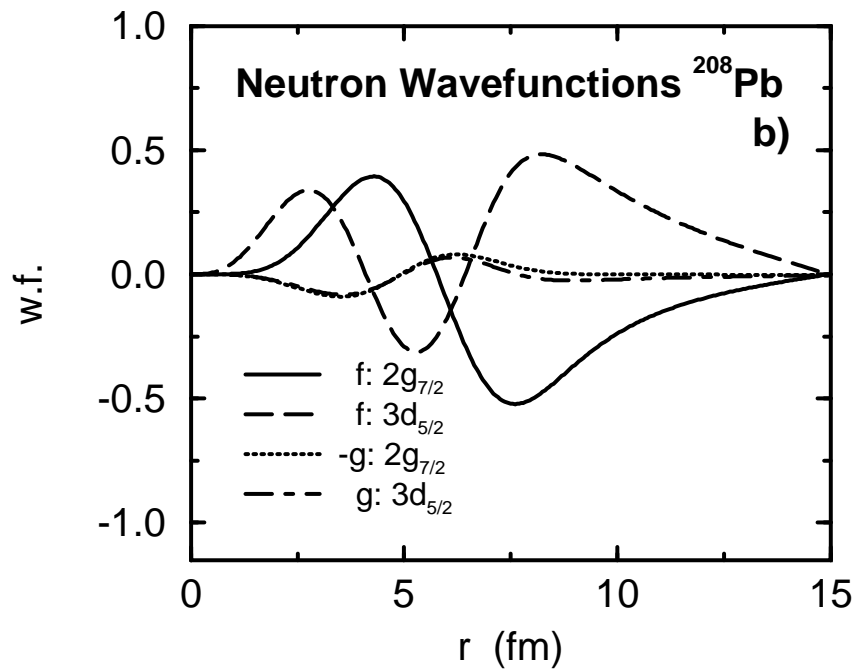
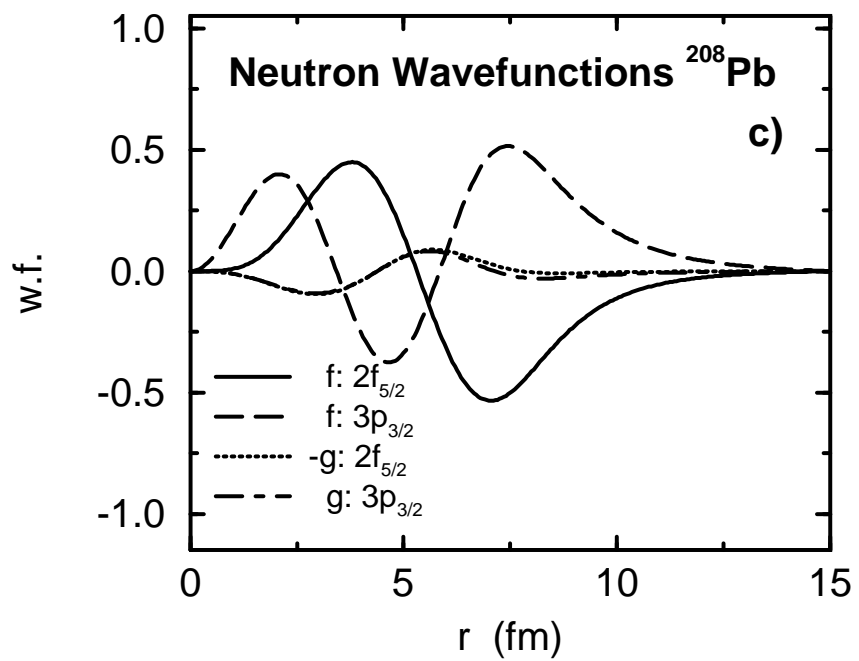
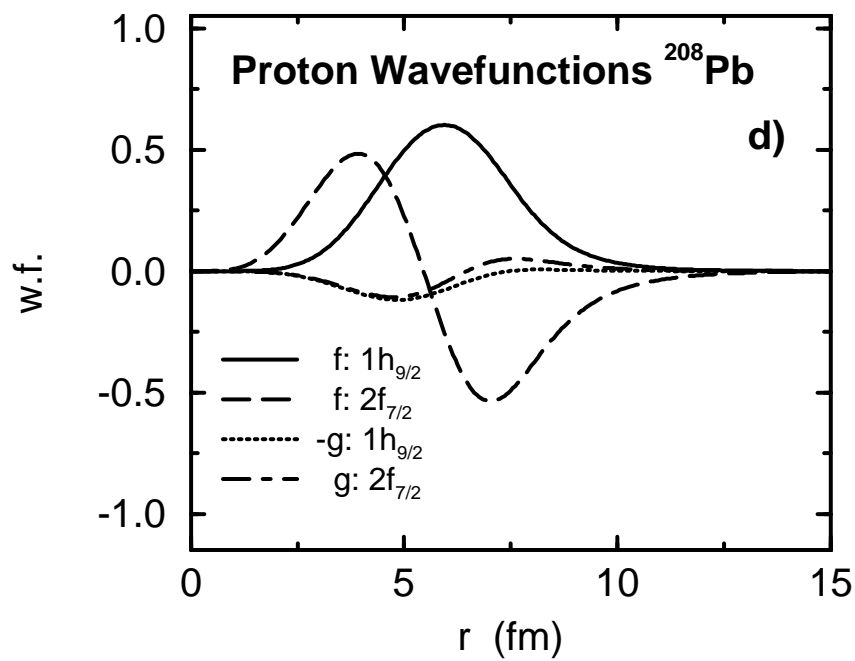
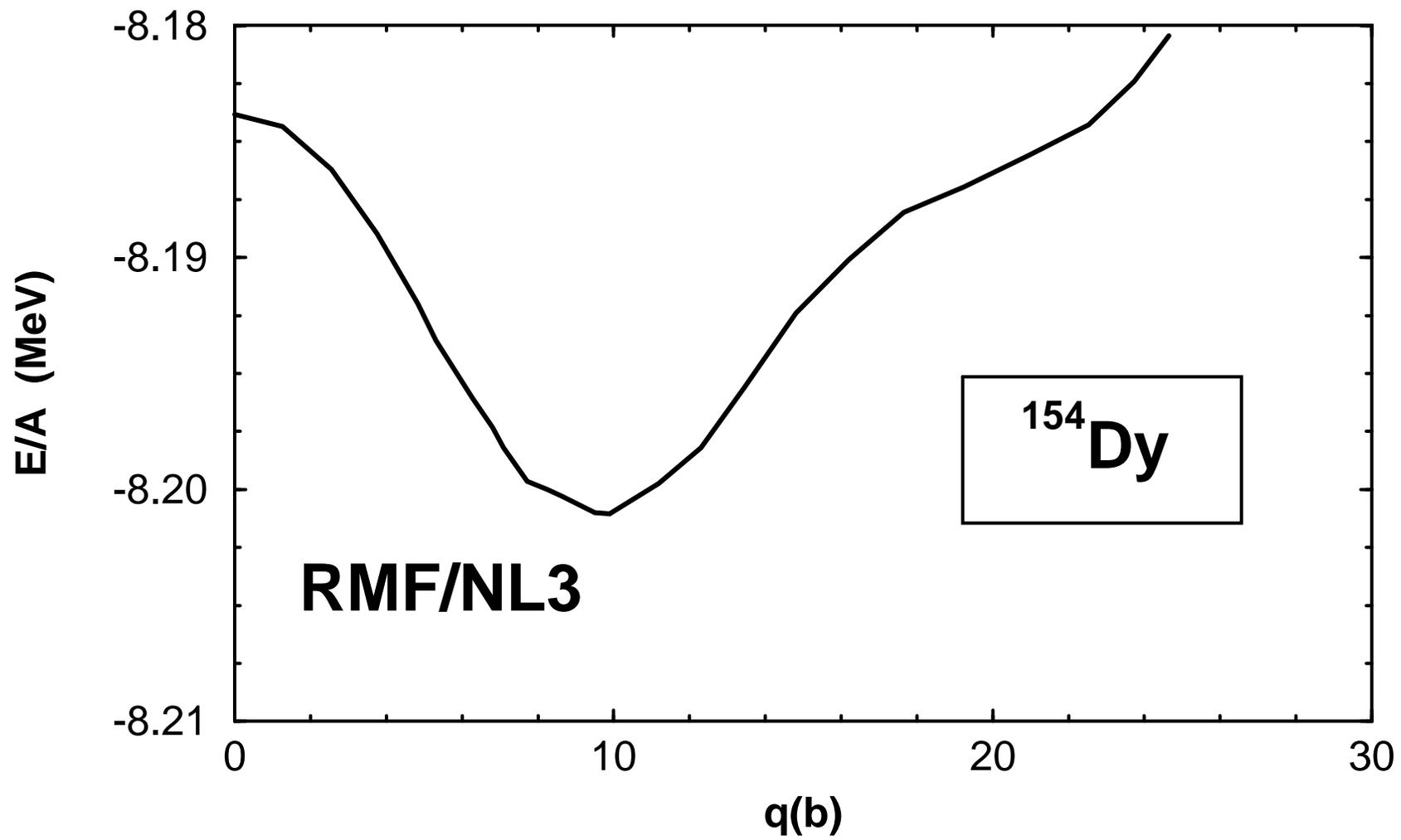
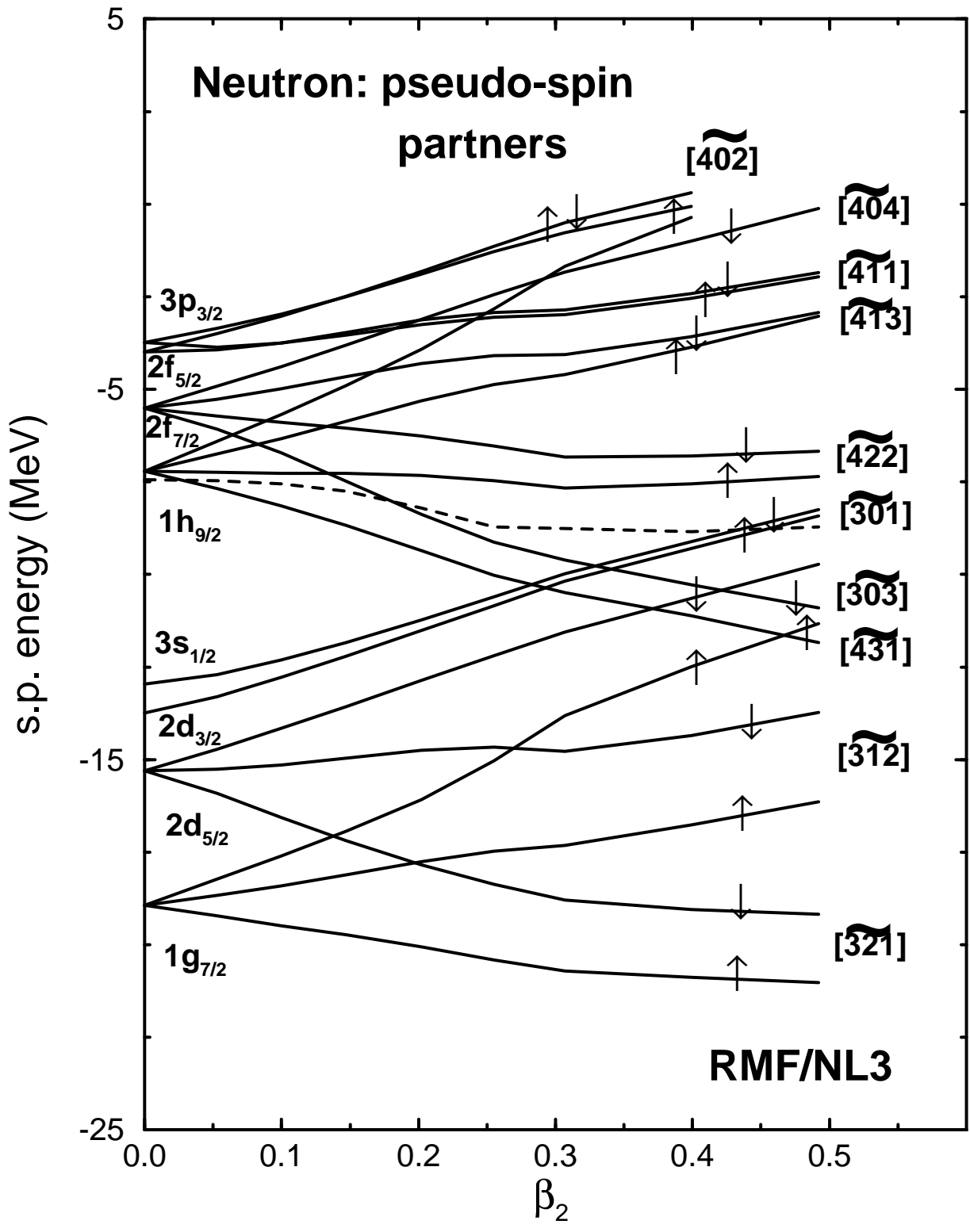
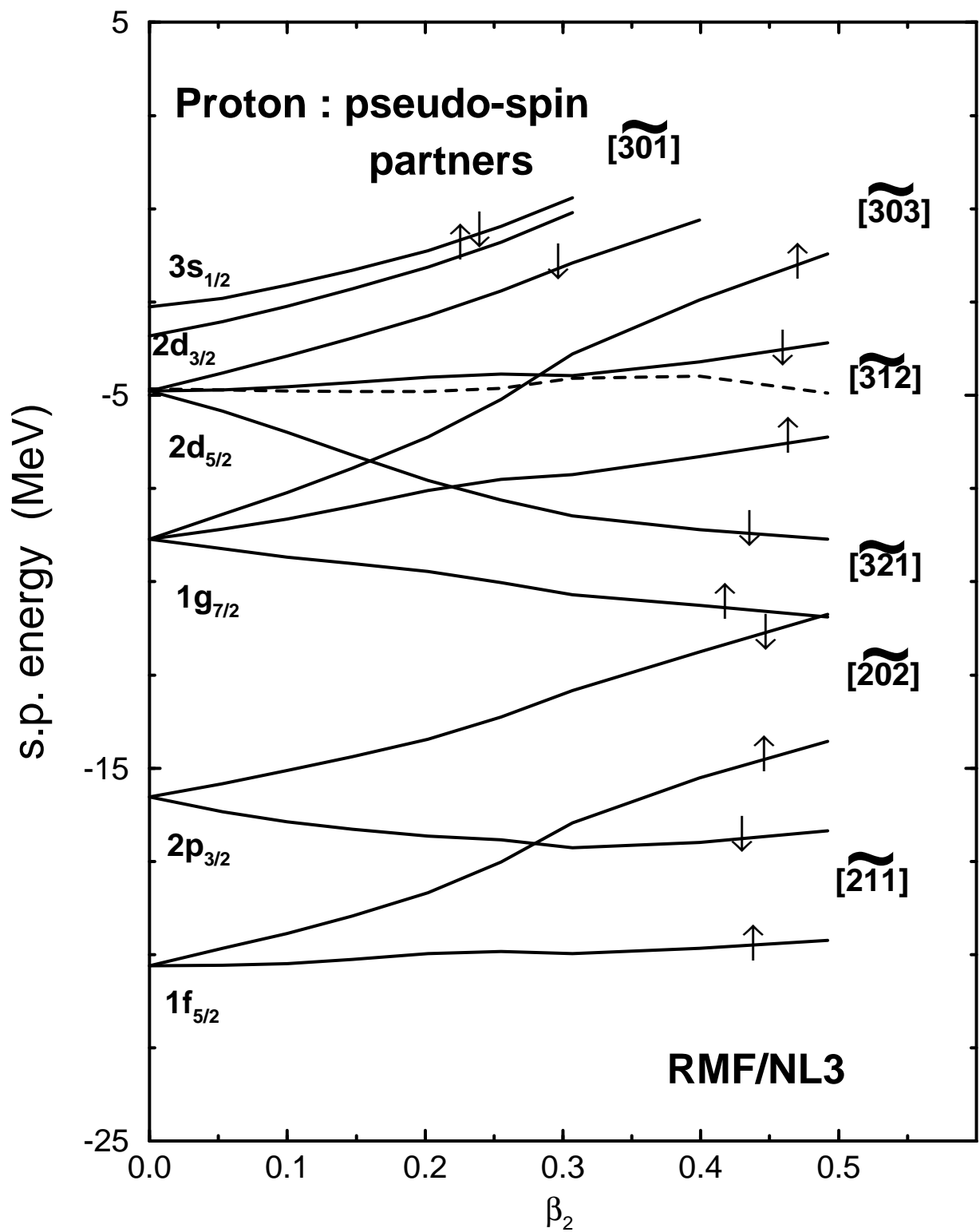
\end{document}